\documentclass{article}
\usepackage{eurosym}
\usepackage{amsfonts}
\usepackage{amsmath}

\newtheorem{theorem}{Theorem}

\newtheorem{corollary}[theorem]{Corollary}

\newtheorem{lemma}[theorem]{Lemma}

\newtheorem{remark}[theorem]{Remark}

\begin{document}

\begin{center}
Novel computational formulas and relations for special numbers and
functions: approach to analysis of Debye functions

Yilmaz Simsek

Akdeniz University Faculty of Science Department of Mathematics 07058
Antalya-Turkey

Email: ysimsek@akdeniz.edu.tr

\bigskip
\end{center}

\textbf{Abstract.} The first goal of this paper is to give a new family of
special function related to the Debye functions. By using these function and
generating function method, we derive many novel formulas and relations
involving the Debye functions, multiple Hurwitz Lerch zeta function which
interpolate the Apostol-Bernoulli numbers of higher order at negative
integers. Proof method of theorems in this paper are different from those of
theory of analytic numbers. The second goal of this paper is to show that
special values and integral presentations of this function are closely
related to unification of the Debye functions, the moment generating
function for the negative binom distribution, the generating functions for
the Apostol-type numbers of higher-order, and the Frobenius-Euler numbers,
the Bernoulli numbers, and the Stirling numbers etc. Moreover, we give
recurrence relations, Maclaurin's series expansion, and approximation
formula for unification of the Debye functions. Finally, we give further
remarks, observations and applications in mathematical physics on the
results of this paper.



\section{Introduction}

The Debye functions have many applications in physical problems, such as the
evaluation of the heat capacity of solids, thermodynamic problems in the
context of crystallographic structure and radiation (\textit{cf}. \cite{Abro}%
, \cite{Cox}, \cite{Debey}, \cite{Engeln}, \cite{Epap}, \cite{Gonzalez}, 
\cite{Grad}, \cite{Guagliardi}, \cite{Guagliardi}, \cite{Guseinov}, \cite%
{Hall}, \cite{Harrison}, \cite{Jordan}, \cite{Khishchenko}, \cite{Macleod}, 
\cite{Ng}, \cite{Passler}). The aim of this paper many novel formulas and
relations for the Debye functions using generating function, multiple
Hurwitz Lerch zeta function, some certain family of special numbers, moment
generating function for the negative binom distribution, and also
approximation formulas involving these function, by the aid of generating
functions methods. Especially, we derive some new formulas for the
Apostol-Bernoulli numbers of higher order and other special finite sums
formulas with the aid of the unification Debye functions method. These
formulas and relations can be used not only in mathematics, but also in some
areas of physics and chemistry.

In order to arrive the motivation of this paper, we use the following
function:%
\begin{equation}
f_{n}(t;\beta ,\theta ,\alpha )=\frac{t^{n}}{\left( \theta e^{t}-\alpha
\right) ^{\beta }},  \label{f1}
\end{equation}%
where $n$, $\alpha $, $\beta $, and $\theta $ are real or complex
parameters. Throughout of this paper, for $w^{\alpha }=e^{\alpha \ln w}$,
assuming that $\ln w$ denotes the principal branch of the multi-valued
function $\ln w$ with the imaginary part $\operatorname{Im}(\ln w)$ constrained by
the interval $(-\pi ,\pi ]$. For instance, for $w\in \mathbb{C}$, one has%
\begin{equation*}
\ln w=\ln \left\vert w\right\vert +i\arg \left( w\right)
\end{equation*}%
where $-\pi <i\arg \left( w\right) \leq \pi .$

We show that the function $f_{n}(t;\beta ,\theta ,\alpha )$ is closely
related to many well-known generating functions for certain family of
special numbers, special functions, integral and series presentations. In
the next sections, we give these representations with palliations of the
function $f_{n}(t;\beta ,\theta ,\alpha )$. We also show that these
presentations involving the Debye functions, the moment generating function
for the negative binom distribution, the multiple Hurwitz Lerch zeta
function, the Riemann zeta function, the generating functions for the
Apostol-type numbers of higher-order, the Frobenius-Euler numbers, the
Bernoulli numbers, and the Stirling numbers, etc.

Some properties of the function $f_{n}(t;\beta ,\theta ,\alpha )$ give as
follows:

The function $f_{n}(t;\beta ,\theta ,\alpha )$ is a meromorf function whose
poles are given by%
\begin{equation*}
t=2\pi ik-\ln \left( \frac{\theta }{\alpha }\right) ,
\end{equation*}%
where $k\in \mathbb{N}$.

The function%
\begin{equation*}
\frac{\alpha ^{\beta }}{t^{n}}f_{n}(t;\beta ,\theta ,\alpha )=\frac{1}{%
\left( \frac{\theta }{\alpha }e^{t}-1\right) ^{\beta }}
\end{equation*}%
gives meromorphic continuation for the generating function of the certain
family of the rational functions association and the $c$-deformed negative
polylogarithms. This function have many applications in the theory of
special numbers involving Frobenious-Euler numbers, the Apostol- Bernoulli
numbers, the real Heisenberg group, the Hilbert-Polya operator and in the
theory of the Lerch transcendent (cf. \cite{Lagaris}).

When $\beta =n$ ($n\in \mathbb{N}$) and $\alpha =1$, we have%
\begin{equation*}
f_{n}(t;n,\theta ,1)=\frac{t^{n}}{\left( \theta e^{t}-1\right) ^{n}}.
\end{equation*}%
This function denotes generating function for the Apostol-Bernoulli numbers
of order $n$, see for detail \cite{SimsekBook2025}-\cite{SrivatavaChoi}.

When $\beta =n$ ($n\in \mathbb{N}$) and $\alpha =\theta =1$, we have%
\begin{equation*}
f_{n}(t;n,1,1)=\frac{t^{n}}{\left( e^{t}-1\right) ^{n}}.
\end{equation*}%
This function denotes denotes generating function for the Bernoulli numbers
of order $n$, see for detail \cite{SimsekBook2025}-\cite{SrivatavaChoi}.

When $\beta =n$ ($n\in \mathbb{N}$) and $\alpha =-1$, we have%
\begin{equation*}
\frac{2^{n}}{t^{n}}f_{n}(t;n,\theta ,-1)=\frac{2^{n}}{\left( \theta
e^{t}+1\right) ^{n}}.
\end{equation*}%
This function denotes denotes generating function for the Apostol-Euler
numbers of order $n$, see for detail \cite{SimsekBook2025}-\cite%
{SrivatavaChoi}.

When $\beta =n$, $\alpha =-1$ and $\theta =1$, we have%
\begin{equation*}
\frac{2^{n}}{t^{n}}f_{n}(t;n,1,-1)=\frac{2^{n}}{\left( e^{t}+1\right) ^{n}}.
\end{equation*}%
This function denotes denotes generating function for the Euler numbers of
order $n$ , see for detail \cite{SimsekBook2025}-\cite{SrivatavaChoi}.

When $\beta =n$ and $\alpha =1$, we have%
\begin{equation*}
2^{n}f_{n}(t;n,\theta ,1)=\frac{2^{n}t^{n}}{\left( \theta e^{t}-1\right) ^{n}%
}.
\end{equation*}%
This function denotes denotes generating function for the Apostol-Genocchi
numbers of order $n$ , see for detail \cite{SimsekBook2025}-\cite%
{SrivatavaChoi}.

When $\beta =n$ and $\alpha =\theta =1$, we have%
\begin{equation*}
2^{n}f_{n}(t;n,1,1)=\frac{2^{n}t^{n}}{\left( e^{t}-1\right) ^{n}}.
\end{equation*}%
This function denotes denotes generating function for the Genocchi numbers
of order $n$, see for detail \cite{SimsekBook2025}-\cite{SrivatavaChoi}.

When $\beta =2n$, $\alpha =2$ and $\theta =1$, we have%
\begin{equation*}
\frac{2^{n}}{t^{2n}}f_{n}(t;2n,1,2)=\frac{2^{n}}{\left( e^{t}-2\right) ^{2n}}%
.
\end{equation*}%
This function denotes denotes generating function for the Fubini numbers of
order $n$, see for details \cite{Kilar}.

The function $f_{n}(t;\beta ,\theta ,\alpha )$ can be written as in terms of 
$\sinh (x)$:%
\begin{equation*}
f_{n}(t;\beta ,\theta ,\alpha )=\frac{t^{n}}{\alpha ^{\beta }e^{\frac{\beta 
}{2}\left( \ln \left( \frac{\theta }{\alpha }\right) +t\right) }\sinh
^{\beta }\left( \frac{\ln \left( \frac{\theta }{\alpha }\right) +t}{2}%
\right) }
\end{equation*}

The function $f_{n}(t;\beta ,\theta ,\alpha )$ is related to the moment
generating function for the negative binom distribution:%
\begin{equation*}
M_{x}\left( w;\frac{\alpha -\theta }{\alpha },\frac{\theta }{\alpha },\beta
\right) =\left( \frac{\frac{\alpha -\theta }{\alpha }}{1-\frac{\theta }{%
\alpha }\exp (w)}\right) ^{\beta },
\end{equation*}%
where $\alpha $, $\beta $, and $\theta $ are positive real parameters such
that $0<\frac{\theta }{\alpha }<1$ and $0<\frac{\alpha -\theta }{\alpha }<1$
with $\frac{\theta }{\alpha }+\frac{\alpha -\theta }{\alpha }=1$ and
generating function for the Apostol-type numbers of order $\beta $. We show
that%
\begin{equation*}
f_{n}(t;\beta ,\theta ,\alpha )=\frac{t^{n-\beta }}{\left( -\alpha ^{\beta
}\right) }M_{x}\left( w;\frac{\alpha -\theta }{\alpha },\frac{\theta }{%
\alpha },\beta \right) .
\end{equation*}%
By applying Theorem 2 and Theorem 3, given in \cite{BRevista}, it is easy to
show that%
\begin{eqnarray*}
f_{n}(t;\beta ,\theta ,\alpha ) &=&\frac{\left( \alpha -\theta \right)
^{\beta }}{\alpha ^{2\beta }}\sum\limits_{v=0}^{\infty }\frac{\mathcal{B}%
_{v+\beta }^{\left( \beta \right) }\left( \frac{\theta }{\alpha }\right) }{%
v!\beta !\binom{\beta +v}{\beta }}t^{v+n-\beta } \\
&=&\frac{1}{\left( -\alpha \right) ^{\beta }}\sum\limits_{v=0}^{\infty }%
\frac{\mu _{v-\beta }\left( \frac{\alpha -\theta }{\alpha },\frac{\theta }{%
\alpha },\beta \right) }{v!}t^{v+n-\beta },
\end{eqnarray*}%
where $\mathcal{B}_{v+\beta }^{\left( \beta \right) }\left( \frac{\theta }{%
\alpha }\right) $ and $\mu _{v-\beta }\left( \frac{\alpha -\theta }{\alpha },%
\frac{\theta }{\alpha },\beta \right) $ denote the Apostol- numbers of order 
$\beta $ and nth moment of the $X$ Negative Binomial random variable with
parameter $\frac{\alpha -\theta }{\alpha }$ and $\beta $, respectively,
defined by%
\begin{equation}
f_{\beta }(t;\beta ,\theta ,1)=\left( \frac{t}{\theta e^{t}-1}\right)
^{\beta }=\sum\limits_{v=0}^{\infty }\mathcal{B}_{v}^{\left( \beta \right)
}\left( \theta \right) \frac{t^{v}}{v!},  \label{abp-h}
\end{equation}%
where $\left\vert t\right\vert <2\pi $ when $\theta =1$; $\left\vert
t\right\vert <\left\vert \ln \left( \theta \right) \right\vert $ when $%
\theta \neq 1$ (\textit{cf}. \cite{SimsekBook2025}-\cite{SimsekBookSpringer1}%
, \cite{SrivatavaChoi}) and%
\begin{equation*}
\mu _{v-\beta }\left( \frac{\alpha -\theta }{\alpha },\frac{\theta }{\alpha }%
,\beta \right) =\frac{d^{v-\beta }}{dt^{v-\beta }}M_{x}\left( t;\frac{\alpha
-\theta }{\alpha },\frac{\theta }{\alpha },\beta \right)
\end{equation*}%
(\textit{cf}. \cite{Bertsekas}, \cite{BRevista}).

Integrate (\ref{f1}), with respect to $t$ from $0$ to $x$, we define the
following function, which is related to the Debye function for special
values of the parameters $\beta ,\theta ,\alpha $:%
\begin{equation}
S_{n}(x;\beta ,\theta ,\alpha )=\frac{n}{x^{n}}\int\limits_{0}^{x}f_{n}(t;%
\beta ,\theta ,\alpha )dt,  \label{100}
\end{equation}%
where $\left\vert t\right\vert <2\pi $ when $\alpha =\theta =1$; $\left\vert
t\right\vert <\left\vert \ln \left( \frac{\theta }{\alpha }\right)
\right\vert $ when $\frac{\theta }{\alpha }\neq 1$\ and $\operatorname{Re}(n)\geq 1$%
. $\ln w$\ ($\left\vert \arg (w)\right\vert <1$) denotes the principal
branch of the many-valued function $\ln (w)=\ln \left\vert w\right\vert
+i\arg (w)$ ($\left\vert w\right\vert >0$). Thus, the function $%
S_{n}(x;\beta ,\theta ,\alpha )$ can also be given in terms of the
Cauch-Euler derivative operator $\partial _{t}=t\frac{d}{dt}$ as follows:%
\begin{equation}
S_{n}(x;\beta ,\theta ,\alpha )=\int\limits_{0}^{x}\frac{\partial _{t}\left( 
\frac{t}{x}\right) ^{n}}{\left( \theta e^{t}-\alpha \right) ^{\beta }}dt.
\label{1}
\end{equation}

It is easy to see that%
\begin{equation*}
\lim_{x\rightarrow 0}S_{n}(x;\beta ,\theta ,\alpha )=0
\end{equation*}%
if $\theta \neq 1$. If $\beta =\theta =\alpha =1$, then%
\begin{equation*}
\lim_{x\rightarrow 0}S_{n}(x;1,1,1)=0,
\end{equation*}%
which studies next section.

PDEs for the function $S_{n}(x;\beta ,\theta ,\alpha )$, by using the
Leibniz rule for differentiation of integral containing variable $x$:%
\begin{equation*}
\left( \beta x\right) ^{\beta }\frac{\partial }{\partial x}S_{n}(x;\beta
,\theta ,\alpha )+n\alpha ^{\beta }x^{\beta -1}S_{n}(x;\beta ,\theta ,\alpha
)=f_{\beta }\left( x;\beta ,\frac{\theta }{\alpha },1\right) ,
\end{equation*}%
and%
\begin{eqnarray*}
&&\alpha ^{\beta }x^{\beta +1}\frac{\partial ^{2}}{\partial x^{2}}%
S_{n}(x;\beta ,\theta ,\alpha )-n\left( n+1\right) \alpha ^{\beta }x^{\beta
}S_{n}(x;\beta ,\theta ,\alpha ) \\
&=&-f_{\beta }\left( x;\beta ,\frac{\theta }{\alpha },1\right) \left(
n+\beta \theta e^{x}f_{1}\left( x;1,\frac{\theta }{\alpha },1\right) \right)
.
\end{eqnarray*}%
It is well-known that special values of $\frac{\partial }{\partial x}%
S_{n}(x;\beta ,\theta ,\alpha )$ and $\frac{\partial ^{2}}{\partial x^{2}}%
S_{n}(x;\beta ,\theta ,\alpha )$\ have many applications in the entropy and
the isochoric heat capacity of solids. Some of these applications have been
studied by Khishchenko \cite{Khishchenko}.

The motivation of this paper is investigate fundamental properties \ and
novel formulas for \ the function $S_{n}(x;\beta ,\theta ,\alpha )$ with the
aid of the function $f(t;\beta ,\theta ,\alpha )$.

Substituting $\theta =\beta =1$ into (\ref{1}), we have the Debye function%
\begin{equation*}
d_{n}(x):=S_{n}(x;1,1,1)=\int\limits_{0}^{x}\frac{\partial _{t}\left( \frac{t%
}{x}\right) ^{n}}{e^{t}-1}dt,
\end{equation*}%
which is certain family of special mathematical functions. The function $%
d_{n}(x)$\ was defined by Dutch physicist Peter Debye on 1912 for $n=3$. The
function $d_{n}(x)$ has many applications in not only solid-state physics
and statistical mechanics when evaluating the energy, heat capacity, and
thermal properties of crystalline solids with the aid of the the Debye
Model, but also heat capacity of solids, thermal conduction, thermal
expansion, $X$-ray and Neutron fiffraction, theory of special functions etc.
The function $d_{n}(x)$\ provides main bridge amongs microscopic atomic
vibrations or phonons and macroscopic thermodynamic values, etc. The
function $d_{n}(x)$ has also many applications in thermodynamic problems
such as in the context of not only crystallographic structure, but also
radiation (\textit{cf}. \cite{Abro}, \cite{Cox}, \cite{Debey}, \cite{Engeln}%
, \cite{Epap}, \cite{Gonzalez}, \cite{Grad}, \cite{Guagliardi}, \cite%
{Guagliardi}, \cite{Guseinov}, \cite{Hall}, \cite{Harrison}, \cite{Jordan}, 
\cite{Khishchenko}, \cite{Macleod}, \cite{Ng}, \cite{Passler}).

For $x=\frac{\phi }{T}$, $\phi $ and $T$\ denote the Debye temperature and
the absolute temperature, respectively, Harrison and Neighboursthe \cite%
{Harrison} gave many significant figures, the tabulated values for the
function $\frac{x}{n}d_{n}(x)$. They gave some approached of the function $%
d_{n}(x)$ for $x>30$ when $x\rightarrow \infty $.

Substituting $\theta =\beta =1$ into (\ref{1}), $S_{n}(x;1,1,\alpha )$ was
studied in Gonzalez \cite{Gonzalez}. The function\ $S_{n}(x;1,1,\alpha )$ is
a generalization of the Debye function $d_{n}(x)$ the parameter $\alpha $
and $\left\vert \alpha \right\vert <1$\ (\textit{cf}. \cite{Gonzalez}).

Substituting $\theta =\alpha =1$ into (\ref{1}), we get $n$-dimensional
Debye functions $S_{n}(x;\beta ,1,1)$, which has many importance
applications in various branches of mathematics, physics, and chemistry (%
\textit{cf}. \cite{Guseinov}).

Ng and Devine \cite{Ng} defined the Debye functions as follows:%
\begin{equation*}
\overline{d}_{p}(x)=\frac{1}{\Gamma (p+1)}\int\limits_{0}^{x}\frac{t^{p}}{%
e^{t}-1}dt,
\end{equation*}%
where $p\geq 1$, $x=a+ib$ with $a>0$ and%
\begin{equation*}
d_{p}(x)=\frac{1}{\Gamma (p+1)}\int\limits_{x}^{\infty }\frac{t^{p}}{e^{t}-1}%
dt.
\end{equation*}%
Due to the above integrals, it is easy to see that%
\begin{equation*}
\zeta (p+1)=d_{p}(x)+\overline{d}_{p}(x)=\sum\limits_{m=1}^{\infty }\frac{1}{%
m^{p+1}}.
\end{equation*}%
The functions $d_{p}(x)$ and $\overline{d}_{p}(x)$\ are known as incomplete
Riemann zeta functions. These functions are also related to the
Bose-Einstein functions. These function can also be used in the solution of
a transcendental equation (\textit{cf}. \cite{Ng}). For $x\in \mathbb{C}$
and $x\in \mathbb{C}$, $\frac{nn!}{x^{n}}\overline{d}_{p}(x)$ approximations
based on the Pade approximation for%
\begin{equation*}
f_{1}(t;1,1,1)=\frac{t}{e^{t}-1}=\sum\limits_{v=0}^{\infty }\frac{B_{v}}{v!}%
t^{v},
\end{equation*}%
where $B_{n}$ denotes the Bernoulli numbers (\textit{cf}. \cite{Harrison}, 
\cite{Luke}, \cite{Ng}).

We summarize next sections of this paper as follows:

In Section 2, recurrence relation, by the integral by part method, and
Maclaurin's series expansion, by the aid of binomial theorem, of the
function $S_{n}(x;\beta ,\theta ,\alpha )$ are given. By using some special
values of the Maclaurin's series expansion of the function $S_{n}(x;\beta
,\theta ,\alpha )$, we show that the function $S_{n}(x;\beta ,\theta ,\alpha
)$ related to many special numbers and polynomials, multiple Hurwitz Lerch
zeta function, the Riemann zeta function, the Debye functions involving the
Apostol-Bernoulli numbers and the Bernoulli numbers. We give remarks and
observations on our results and their special cases related to references.

In Section 2, approximation formula, integral and series presentations for $%
\lim_{x\rightarrow \infty }\frac{x^{n}S_{n}(x;\beta ,\theta ,\alpha )}{n}$
are given. Some computational formulas for $\lim_{x\rightarrow \infty }\frac{%
x^{n}S_{n}(x;\beta ,\theta ,\alpha )}{n}$ and $\lim_{n\rightarrow
0}S_{n}(x;\beta ,\theta ,\alpha )$ are also given. We also give remarks and
observations on our results and their special cases related to references..
Finally, we give conclusion section.

\section{Recurrence relation and Maclaurin's series expansion of the
function $S_{n}(x;\protect\beta ,\protect\theta ,\protect\alpha )$}

In this section, we give recurrence relation and Maclaurin's series
expansion of the function $S_{n}(x;\beta ,\theta ,\alpha )$. We also give
some special values of the Maclaurin's series expansion of the function $%
S_{n}(x;\beta ,\theta ,\alpha )$.

In order to give recurrence relation for the function $S_{n}(x;\beta ,\theta
,\alpha )$, we assume that $n$ and $\beta $ are integers. By applying
partial integration to equation(\ref{1}) taking $u=\frac{t^{n}}{\left(
\theta e^{t}-\alpha \right) ^{\beta }}$ and $dt=dv$, after some
calculations, we get the following theorem:

\begin{theorem}
(Recurrence relation for the function $S_{n}(x;\beta ,\theta ,\alpha )$)%
\begin{eqnarray}
(1+n)^{2}S_{n}(x;\beta ,\theta ,\alpha ) &=&nx\beta \left( S_{n+1}(x;\beta
,\theta ,\alpha )+\alpha S_{n+1}(x;\beta +1,\theta ,\alpha )\right)  \notag
\\
&&+\frac{n\left( n+1\right) x}{\left( \theta e^{x}-\alpha \right) ^{\beta }}.
\label{rr}
\end{eqnarray}
\end{theorem}

By using equation (\ref{rr}), not only the function $S_{n}(x;\beta ,\theta
,\alpha )$, but also for special values of the parameters\ $\beta ,\theta
,\alpha $,\ the Debye functions can be easily evaluated.

Substituting $\theta =\alpha =1$ into (\ref{rr}), we obtain 
\begin{eqnarray*}
(1+n)^{2}S_{n}(x;\beta ,1,1) &=&nx\beta \left( S_{n+1}(x;\beta ,1,1)+\alpha
S_{n+1}(x;\beta +1,1,1)\right) \\
&&+\frac{n\left( n+1\right) x}{\left( e^{x}-1\right) ^{\beta }}
\end{eqnarray*}%
(\textit{cf}. \cite{Guseinov}).

By combining (\ref{1}) with following generating function for the
Apostol-type Frobenius-Euler numbers of order $\beta $.%
\begin{equation}
\frac{\left( 1-\alpha \right) ^{\beta }}{t^{n}}f_{n}(t;\beta ,\theta ,\alpha
)=\sum\limits_{v=0}^{\infty }\mathcal{H}_{v}^{\left( \beta \right) }\left(
\theta ,\alpha \right) \frac{t^{v}}{v!},  \label{AFE}
\end{equation}%
where $\alpha \in 
\mathbb{C}
$ (\textit{cf}. \cite{SimsekBook2025}-\cite{SimsekBookSpringer1}, \cite%
{SrivatavaChoi}), we get%
\begin{equation*}
S_{n}(x;\beta ,\theta ,\alpha )=\frac{n}{\left( 1-\alpha \right) ^{\beta
}x^{n}}\sum\limits_{v=0}^{\infty }\frac{\mathcal{H}_{v}^{\left( \beta
\right) }\left( \theta ,\alpha \right) }{v!}\int\limits_{0}^{x}t^{n+v}dt.
\end{equation*}%
Therefore, \ the Maclaurin's series expansion of the function $S_{n}(x;\beta
,\theta ,\alpha )$ can be given by the following theorem:

\begin{theorem}
(Maclaurin's series expansion of the function $S_{n}(x;\beta ,\theta ,\alpha
)$)%
\begin{equation}
S_{n}(x;\beta ,\theta ,\alpha )=\frac{n}{\left( 1-\alpha \right) ^{\beta }}%
\sum\limits_{v=0}^{\infty }\frac{\mathcal{H}_{v}^{\left( \beta \right)
}\left( \theta ,\alpha \right) }{\left( n+v+1\right) v!}x^{v+1}.  \label{1a}
\end{equation}
\end{theorem}

By combining (\ref{1}) with (\ref{abp-h}), we get the following Maclaurin's
series expansion of the function $S_{n}(x;\beta ,\theta ,\alpha )$ for $%
\left\vert x\right\vert <2\pi $ when $\alpha =\theta =1$; $\left\vert
t\right\vert <\left\vert \ln \left( \frac{\theta }{\alpha }\right)
\right\vert $ when $\frac{\theta }{\alpha }\neq 1$\ and $\operatorname{Re}(n)\geq 1$:%
\begin{equation}
S_{n}(x;\beta ,\theta ,\alpha )=\frac{n}{\alpha ^{\beta }x^{n}}%
\sum\limits_{c=0}^{\infty }\frac{\mathcal{B}_{c}^{\left( \beta \right)
}\left( \frac{\theta }{\alpha }\right) }{c!}\int\limits_{0}^{x}t^{c+n-\beta
}dt.  \label{1ABN0}
\end{equation}%
Thus, we have the following theorem:

\begin{theorem}
(Maclaurin's series expansion of the function $S_{n}(x;\beta ,\theta ,\alpha
)$)%
\begin{equation}
S_{n}(x;\beta ,\theta ,\alpha )=\frac{n}{\alpha ^{\beta }}%
\sum\limits_{c=0}^{\infty }\frac{\mathcal{B}_{c}^{\left( \beta \right)
}\left( \frac{\theta }{\alpha }\right) }{\left( c+n-\beta +1\right) c!}%
x^{c-\beta +1},  \label{1b}
\end{equation}%
where $\alpha \neq 0$, $\left\vert x\right\vert <2\pi $ when $\alpha =\theta
=1$; $\left\vert t\right\vert <\left\vert \ln \left( \frac{\theta }{\alpha }%
\right) \right\vert $ when $\frac{\theta }{\alpha }\neq 1$\ and $\operatorname{Re}%
(n)\geq 1$.
\end{theorem}

Some special equation (\ref{1b}) are given as follows:

Combining (\ref{1b}) with (\ref{rr}), we get%
\begin{eqnarray*}
&&(1+n)\sum\limits_{c=0}^{\infty }\frac{\mathcal{B}_{c}^{\left( \beta
\right) }\left( \frac{\theta }{\alpha }\right) }{\left( c+n-\beta +1\right)
c!}x^{c-\beta -1}-\alpha ^{\beta }\sum\limits_{c=0}^{\infty }\frac{\mathcal{B%
}_{c}^{\left( \beta \right) }\left( \frac{\theta }{\alpha }\right) }{c!}x^{c}
\\
&=&\beta \sum\limits_{c=0}^{\infty }\frac{\mathcal{B}_{c}^{\left( \beta
\right) }\left( \frac{\theta }{\alpha }\right) }{\left( c+n-\beta +2\right)
c!}x^{c-\beta +2}+\beta \sum\limits_{c=0}^{\infty }\frac{\mathcal{B}%
_{c}^{\left( \beta +1\right) }\left( \frac{\theta }{\alpha }\right) }{\left(
c+n-\beta +3\right) c!}x^{c-\beta +3}.
\end{eqnarray*}

Assuming that $\beta $ and $n$ are nonnegative integers, after some
calculations the above equation, we get%
\begin{eqnarray*}
&&(1+n)\sum\limits_{c=0}^{\infty }\frac{\mathcal{B}_{c+\beta -1}^{\left(
\beta \right) }\left( \frac{\theta }{\alpha }\right) }{\binom{c}{\beta -1}%
\left( \beta -1\right) !\left( c+n\right) c!}x^{c}-\alpha ^{\beta
}\sum\limits_{c=0}^{\infty }\frac{\mathcal{B}_{c}^{\left( \beta \right)
}\left( \frac{\theta }{\alpha }\right) }{c!}x^{c} \\
&=&\beta \sum\limits_{c=0}^{\infty }\frac{\mathcal{B}_{c+\beta -2}^{\left(
\beta \right) }\left( \frac{\theta }{\alpha }\right) }{\binom{c}{\beta -2}%
\left( \beta -2\right) !\left( c+n\right) c!}x^{c}+\beta
\sum\limits_{c=0}^{\infty }\frac{\mathcal{B}_{c+\beta -3}^{\left( \beta
+1\right) }\left( \frac{\theta }{\alpha }\right) }{\binom{c}{\beta -3}\left(
\beta -3\right) !\left( c+n\right) c!}x^{c}.
\end{eqnarray*}%
Now, comparing the coefficients of $\frac{x^{c}}{c!}$ on both sides of the
above equation, after some calculations, we get the following recurrence
relation for $\mathcal{B}_{c}^{\left( \beta \right) }\left( \frac{\theta }{%
\alpha }\right) $:

\begin{theorem}
Let $n\in \mathbb{N}_{0}$ and $c+\beta \geq 3$. Then%
\begin{eqnarray}
&&\frac{(1+n)\mathcal{B}_{c+\beta -1}^{\left( \beta \right) }\left( \frac{%
\theta }{\alpha }\right) }{\binom{c}{\beta -1}\left( \beta -1\right) !}-%
\frac{\beta \mathcal{B}_{c+\beta -2}^{\left( \beta \right) }\left( \frac{%
\theta }{\alpha }\right) }{\binom{c}{\beta -2}\left( \beta -2\right) !}-%
\frac{\beta \mathcal{B}_{c+\beta -3}^{\left( \beta +1\right) }\left( \frac{%
\theta }{\alpha }\right) }{\binom{c}{\beta -3}\left( \beta -3\right) !}
\label{arr} \\
&=&\left( c+n\right) \alpha ^{\beta }\mathcal{B}_{c}^{\left( \beta \right)
}\left( \frac{\theta }{\alpha }\right) .  \notag
\end{eqnarray}
\end{theorem}

Putting $c=\beta $ and $\alpha =1$\ in (\ref{arr}), we get the following
result:

\begin{corollary}
Let $n\in \mathbb{N}_{0}$ and $2c\geq 3$. Then%
\begin{equation}
(1+n)\mathcal{B}_{2c-1}^{\left( c\right) }\left( \theta \right) -c\mathcal{B}%
_{2c-2}^{\left( c\right) }\left( \theta \right) -c\mathcal{B}_{2c-3}^{\left(
c+1\right) }\left( \theta \right) =\left( c+n\right) c!\mathcal{B}%
_{c}^{\left( c\right) }\left( \theta \right) .  \label{rrev}
\end{equation}
\end{corollary}

In order to evaluate of the function $S_{n}(x;\beta ,\theta ,\alpha )$ and (%
\ref{rrev}), we need the following explicit formula for the
Apostol-Bernoulli number of order $\beta $, see \cite[p. 94, Eq. (19)]%
{SrivatavaChoi}:%
\begin{equation*}
\mathcal{B}_{c}^{\left( \beta \right) }\left( \frac{\theta }{\alpha }\right)
=\beta !\binom{c}{\beta }\sum\limits_{d=0}^{c-\beta }\binom{\beta +d-1}{d}%
\frac{d!\left( -\frac{\theta }{\alpha }\right) ^{d}}{\left( \frac{\theta }{%
\alpha }-1\right) ^{d+\beta }}S_{2}(c-\beta ,d),
\end{equation*}%
where $\beta ,c\in \mathbb{N}_{0}$ and $\frac{\theta }{\alpha }\in \mathbb{C}
$, where $S_{2}(c,d)$ denotes the Stirling numbers of the second kind,
defined by%
\begin{equation*}
\frac{1}{d!}\left( e^{u}-1\right) ^{d}=\sum\limits_{c=0}^{\infty }\frac{%
S_{2}(c,d)}{d!}u^{d}.
\end{equation*}

Substituting $\alpha =\theta =1$ into (\ref{1b}), since $\mathcal{B}%
_{0}\left( \frac{\theta }{\alpha }\right) =0$ for $\frac{\theta }{\alpha }%
\neq 1$, we get%
\begin{equation*}
S_{n}(x;1,\theta ,\alpha )=\frac{n}{\alpha }\sum\limits_{c=1}^{\infty }\frac{%
\mathcal{B}_{c}\left( \frac{\theta }{\alpha }\right) }{\left( c+n\right) c!}%
x^{c},
\end{equation*}%
where $\alpha \neq 0$.

Substituting $\alpha =\theta =1$ into (\ref{1b}), we get%
\begin{equation*}
S_{n}(x;\beta ,1,1)=n\sum\limits_{c=0}^{\infty }\frac{B_{c}^{\left( \beta
\right) }}{c!}\frac{x^{c-\beta +1}}{c+n-\beta +1}.
\end{equation*}%
When $\beta =1$, the above formula reduces to the well-known Maclaurin's
series expansion of the function $d_{n}(x)$ for $\left\vert x\right\vert
<2\pi $:%
\begin{eqnarray}
d_{n}(x) &=&n\sum\limits_{v=0}^{\infty }\frac{B_{c}}{c!}\frac{x^{c}}{c+n}
\label{dB} \\
&=&1-\frac{nx}{2\left( n+1\right) }+\sum\limits_{c=2}^{\infty }\frac{B_{2c}}{%
\left( 2c+n\right) \left( 2c\right) !}x^{2c}  \notag
\end{eqnarray}%
since for $c>1$, $B_{2c+1}=0$ (\textit{cf}. \cite{Guseinov}\thinspace\ \cite%
{Gonzalez}, \cite{Ng}).

Simsek \cite{BRevista} showed that $n$th moment of the $X$ Negative Binomial
random variable with parameter $p$ and $r$ has the following formulas:%
\begin{equation}
\mu _{n}(p,q,r)=\frac{(-p)^{r}}{\binom{n+r}{r}r!}\mathcal{B}_{n+r}^{(r)}(q),
\label{nm}
\end{equation}

Combining (\ref{nm}) with (\ref{1b}), assuming that $\beta \in \mathbb{N}$, $%
0<\frac{\theta }{\alpha }<1$\ and $q=\frac{\theta }{\alpha }$ with $p+q=1$,
we obtain the following novel formula:

\begin{theorem}
Let $\theta \neq \alpha $ and $\beta \in \mathbb{N}$. Then%
\begin{equation*}
S_{n}(x;\beta ,\theta ,\alpha )=n\sum\limits_{v=\beta }^{\infty }\binom{v}{%
\beta }\frac{\beta !\mu _{v-\beta }\left( \frac{\alpha -\theta }{\alpha },%
\frac{\theta }{\alpha },\beta \right) }{\left( \theta -\alpha \right)
^{\beta }\left( v+n-\beta +1\right) }\frac{x^{v-\beta +1}}{v!},
\end{equation*}
\end{theorem}

Combining (\ref{1a}) with (\ref{1b}), we get%
\begin{equation}
\sum\limits_{v=0}^{\infty }\frac{\mathcal{B}_{v}^{\left( \beta \right)
}\left( \frac{\theta }{\alpha }\right) }{\left( v+n-\beta +1\right) v!}%
x^{v}=\left( \frac{x\alpha }{1-\alpha }\right) ^{^{\beta
}}\sum\limits_{v=0}^{\infty }\frac{\mathcal{H}_{v}^{\left( \beta \right)
}\left( \theta ,\alpha \right) }{\left( n+v+1\right) v!}x^{v}  \label{1c}
\end{equation}

If $\beta $ is a positive integer, then equation (\ref{1c}) reduces to%
\begin{equation*}
\sum\limits_{v=0}^{\infty }\frac{\mathcal{B}_{v}^{\left( \beta \right)
}\left( \frac{\theta }{\alpha }\right) }{\left( v+n-\beta +1\right) v!}%
x^{v}=\left( \frac{\alpha }{1-\alpha }\right) ^{^{\beta
}}\sum\limits_{v=0}^{\infty }\frac{\binom{n}{\beta }\beta !\mathcal{H}%
_{v-\beta }^{\left( \beta \right) }\left( \theta ,\alpha \right) }{\left(
n+v-\beta +1\right) v!}x^{v}.
\end{equation*}%
By comparing the coefficients of $\frac{x^{v}}{v!}$ on both sides of the
above equation, we obtain the following result:

\begin{theorem}
If $\beta $ is a positive integer, then we have%
\begin{equation*}
\mathcal{B}_{v}^{\left( \beta \right) }\left( \frac{\theta }{\alpha }\right)
=\left( \frac{\alpha }{1-\alpha }\right) ^{^{\beta }}\binom{n}{\beta }\beta !%
\mathcal{H}_{v-\beta }^{\left( \beta \right) }\left( \theta ,\alpha \right) .
\end{equation*}
\end{theorem}

Equation (\ref{1}) can also be given as follows:%
\begin{equation*}
\frac{x^{n}}{n}S_{n}(x;\beta ,\theta ,\alpha )=\int\limits_{0}^{\infty }%
\frac{t^{n}}{\left( \theta e^{t}-\alpha \right) ^{\beta }}%
dt-\int\limits_{x}^{\infty }\frac{t^{n}}{\left( \theta e^{t}-\alpha \right)
^{\beta }}dt.
\end{equation*}%
By applying the binomial theorem with $\left\vert \frac{\alpha }{\theta }%
e^{t}\right\vert <1$, we simplify the above equation by the aid of the
Laplace transform and integrated by parts of the following known integral
formula, given by Gradshteyn and Ryzhik \cite[p. 112]{Grad}:%
\begin{equation}
\int t^{n}e^{ct}dt=e^{ct}\left( \frac{t^{n}}{c}+\sum\limits_{j=1}^{n}(-1)^{j}%
\binom{n}{j}j!\frac{t^{n-j}}{c^{j+1}}\right) ,  \label{if-1}
\end{equation}%
we get%
\begin{eqnarray*}
&&S_{n}(x;\beta ,\theta ,\alpha ) \\
&=&\frac{n\Gamma (n+1)}{\theta ^{\beta }x^{n}}\sum\limits_{v=0}^{\infty }%
\binom{\beta +v-1}{v}\frac{\left( \frac{\alpha }{\theta }\right) ^{v}}{%
(v+\beta )^{n+1}} \\
&&-\frac{n}{\theta ^{\beta }x^{n}}\sum\limits_{v=0}^{\infty }\binom{\beta
+v-1}{v}\left( \frac{\alpha }{\theta }\right) ^{v}\lim_{R\rightarrow \infty
}e^{-t(v+\beta )}\left( \frac{t^{n}}{v+\beta }+\sum\limits_{j=1}^{n}\frac{%
(-1)^{j}\binom{n}{j}j!t^{n-j}}{(v+\beta )^{j+1}}\right) \left\vert
_{x}^{R}\right. .
\end{eqnarray*}%
After some calculations, we get the following Lemma:

\begin{lemma}
Let $\theta \neq 0$ and $n+1>0$. Then%
\begin{eqnarray}
&&S_{n}(x;\beta ,\theta ,\alpha )  \label{X-1} \\
&=&\frac{n\Gamma \left( n+1\right) }{\theta ^{\beta }x^{n}}%
\sum\limits_{v=0}^{\infty }\binom{\beta +v-1}{v}\frac{\left( \frac{\alpha }{%
\theta }\right) ^{v}}{(v+\beta )^{n+1}}  \notag \\
&&-\frac{n}{x^{n}\theta ^{\beta }}\sum\limits_{v=0}^{\infty }\binom{\beta
+v-1}{v}\left( \frac{\alpha }{\theta }\right) ^{v}\frac{e^{-x(v+\beta )}}{%
v+\beta }  \notag \\
&&-\frac{n}{\theta ^{\beta }}\sum\limits_{v=0}^{\infty }\binom{\beta +v-1}{v}%
\left( \frac{\alpha }{\theta }\right) ^{v}e^{-x(v+\beta
)}\sum\limits_{j=1}^{n}\frac{(-1)^{j}\binom{n}{j}j!}{(v+\beta )^{j+1}x^{j}}.
\notag
\end{eqnarray}
\end{lemma}

\begin{remark}
Substituting $\beta =\theta =\alpha =1$ into (\ref{X-1}), we get%
\begin{equation}
S_{n}(x;1,1,1)=\frac{n\Gamma \left( n+1\right) }{x^{n}}\zeta (n+1)-\frac{n}{%
x^{n}}\sum\limits_{v=0}^{\infty }\frac{e^{-x(v+1)}}{v+1}-n\sum%
\limits_{v=0}^{\infty }e^{-x(v+\beta )}\sum\limits_{j=1}^{n}\frac{(-1)^{j}%
\binom{n}{j}j!}{(v+1)^{j+1}x^{j}},  \notag
\end{equation}%
where $x>0$. Putting $n=3$ in the above equation and using the following
well-known series%
\begin{equation*}
\sum\limits_{v=0}^{\infty }\frac{1}{(v+1)^{4}}=\frac{\pi ^{4}}{90},
\end{equation*}%
we have%
\begin{equation*}
S_{3}(x;1,1,1)=\frac{\pi ^{4}}{5x^{3}}-3\sum\limits_{v=0}^{\infty }\frac{%
e^{-x(v+1)}}{v+1}-3\sum\limits_{v=1}^{\infty
}e^{-x(v+1)}\sum\limits_{j=1}^{3}\frac{(-1)^{j}\binom{3}{j}j!}{%
(v+1)^{j+1}x^{j}},
\end{equation*}%
which is similar to equation (17) in \cite{Khishchenko}.
\end{remark}

Combining (\ref{X-1}) with equation (\ref{1b}), we get the next theorem,
which is related to the multiple Hurwitz Lerch zeta function, defined by%
\begin{eqnarray*}
\Phi _{m}\left( w,z;x\right) &=&\sum\limits_{v=0}^{\infty }\binom{v+m-1}{v}%
\frac{w^{v}}{(v+x)^{s}} \\
&=&\sum\limits_{v=0}^{\infty }\binom{v+m-1}{m-1}\frac{w^{v}}{(v+x)^{s}},
\end{eqnarray*}%
where $m\in \mathbb{N}$, $s\in \mathbb{C}$ with $s=a+ib$, $x\in \mathbb{C}%
\setminus \left\{ 0,-1,-2,-3,\ldots \right\} $; $s\in \mathbb{C}$ when $%
\left\vert w\right\vert <1$; $a>m$ when $\left\vert w\right\vert =1$ (cf. 
\cite[p. 206, Eq. (16)]{SrivatavaChoi}). When $m=1$, $\Phi _{1}\left(
w,z;x\right) $\ reduces to the Lerch transcendent, which satisfies not only
the differential-difference equations%
\begin{equation*}
\left( w\frac{\partial }{\partial w}+x\right) \Phi _{1}\left( w,z;x\right)
=\Phi _{1}\left( w,z-1;x\right)
\end{equation*}%
and%
\begin{equation*}
\frac{\partial }{\partial x}\Phi _{1}\left( w,z;x\right) =-z\Phi _{1}\left(
w,z+1;x\right) ,
\end{equation*}%
but also the linear partial differential equation:%
\begin{equation*}
\left( w\frac{\partial }{\partial w}+x\right) \frac{\partial }{\partial x}%
\Phi _{1}\left( w,z;x\right) =-w\Phi _{1}\left( w,z;x\right)
\end{equation*}%
(cf. \cite{Lagaris}).

\begin{theorem}
Let $\beta $ be a positive integer. Then%
\begin{eqnarray}
&&\sum\limits_{v=0}^{\infty }\frac{\mathcal{B}_{v}^{\left( \beta \right)
}\left( \frac{\theta }{\alpha }\right) }{\left( v+n-\beta +1\right) v!}%
x^{v-\beta +1}  \label{XT-1} \\
&=&\frac{\Gamma \left( n+1\right) }{n}\Phi _{\beta }\left( \frac{\alpha }{%
\theta },n+1;\beta \right) -e^{-x\beta }\Phi _{\beta }\left( \frac{\alpha }{%
\theta }e^{-x},1;\beta \right)  \notag \\
&&-\frac{1}{n}\sum\limits_{j=1}^{n}(-1)^{j}\binom{n}{j}j!x^{n-j}e^{-x\beta
}\Phi _{\beta }\left( \frac{\alpha }{\theta }e^{-x},j+1;\beta \right) . 
\notag
\end{eqnarray}
\end{theorem}

Substituting $\beta =\theta =1$ into (\ref{XT-1}), we get%
\begin{eqnarray}
\sum\limits_{v=0}^{\infty }\frac{\mathcal{B}_{v}\left( \frac{1}{\alpha }%
\right) }{\left( v+n\right) v!}x^{v} &=&\frac{\Gamma \left( n+1\right) }{n}%
\Phi \left( \alpha ,n+1;1\right) -e^{-x}\Phi \left( \alpha e^{-x},1;1\right)
\notag \\
&&-\frac{1}{n}\sum\limits_{j=1}^{n}(-1)^{j}\binom{n}{j}j!x^{n-j}e^{-x}\Phi
\left( \alpha e^{-x},j+1;1\right) .  \notag
\end{eqnarray}%
For $\left\vert \alpha \right\vert <1$, $n>0$, combining the above equation
with the Lerch zeta function, we get the following theorem:

\begin{theorem}
Let $\left\vert \alpha \right\vert <1$, $n>0$. Then%
\begin{eqnarray}
\sum\limits_{v=0}^{\infty }\frac{\mathcal{B}_{v}\left( \frac{1}{\alpha }%
\right) x^{v+n}}{\left( v+n\right) v!} &=&\alpha n\Gamma (n)\Phi (\alpha
,n+1,1)-\frac{1}{e^{x}}\Phi (\alpha e^{-x},1,1)  \label{1aa} \\
&&-\frac{\alpha }{e^{x}}\sum\limits_{j=1}^{n}(-1)^{j}\binom{n}{j}%
j!x^{n-j}\Phi _{1}\left( \alpha e^{-x},j+1;1\right) .  \notag
\end{eqnarray}
\end{theorem}

\section{Approximation formula with the aid of integral and series
presentations for $\lim_{x\rightarrow \infty }\frac{x^{n}S_{n}(x;\protect%
\beta ,\protect\theta ,\protect\alpha )}{n}$}

In this section, we give an approximation formula of the function $%
S_{n}(x;\beta ,\theta ,\alpha )$ by using integral and series presentations
for the function $\lim_{x\rightarrow \infty }\frac{x^{n}S_{n}(x;\beta
,\theta ,\alpha )}{n}$. We show that $\lim_{x\rightarrow \infty }\frac{%
x^{n}S_{n}(x;\beta ,\theta ,\alpha )}{n}$ can be representation in terms
interpolation functions for the Apostol-Bernoulli numbers of order $\beta $.
Some special values of these integral and series presentations are also
given. In order to give some special values of the above limit computations:%
\begin{equation}
\lim_{x\rightarrow \infty }\frac{x^{n}S_{n}(x;\beta ,\theta ,\alpha )}{n}%
=\int\limits_{0}^{\infty }f_{n}(t;\beta ,\theta ,\alpha )dt,  \label{lc}
\end{equation}%
we need to give the following limit values of equation (\ref{1}) and (\ref%
{1b}), for $n,\beta \in \mathbb{N}$, we get%
\begin{equation}
S_{n}(x;\beta ,\theta ,\alpha )=\frac{n}{\alpha ^{\beta }x^{n}}%
\int\limits_{0}^{x}t^{n}\left( \frac{1}{\frac{\theta }{\alpha }e^{t}-1}%
\right) ^{\beta }dt.  \label{1bl}
\end{equation}%
By using (\ref{1ABN0}), equation (\ref{1bl}) reduces to the generating
function for the Apostol-Bernoulli numbers of order $\beta $:%
\begin{equation}
S_{n}(x;\beta ,\theta ,\alpha )=\frac{n}{\alpha ^{\beta }x^{n}}%
\int\limits_{0}^{x}\sum\limits_{c=0}^{\infty }\frac{\mathcal{B}_{c}^{\left(
\beta \right) }\left( \frac{\theta }{\alpha }\right) }{c!}t^{n+c-\beta }dt
\label{1ABN}
\end{equation}%
We integrate term by term of the series, which is uniformly continuous in
the range $\left\vert t\right\vert <2\pi $ when $\frac{\theta }{\alpha }=1$
and $\left\vert t\right\vert <\left\vert \ln \left( \frac{\theta }{\alpha }%
\right) \right\vert $ when $\frac{\theta }{\alpha }\neq 1$. Therefore, we
have%
\begin{eqnarray*}
S_{n}(x;\beta ,\theta ,\alpha ) &=&\frac{n}{\alpha ^{\beta }x^{n}}%
\sum\limits_{c=0}^{\infty }\frac{\mathcal{B}_{c}^{\left( \beta \right)
}\left( \frac{\theta }{\alpha }\right) }{c!}\int\limits_{0}^{x}t^{n+c-\beta
}dt \\
&=&\frac{n}{\alpha ^{\beta }}\sum\limits_{c=0}^{\infty }\frac{\mathcal{B}%
_{c}^{\left( \beta \right) }\left( \frac{\theta }{\alpha }\right) }{%
(n+c+1-\beta )c!}x^{c+1-\beta },
\end{eqnarray*}%
which yields%
\begin{eqnarray*}
&&\lim_{n\rightarrow 0}S_{n}(x;\beta ,\theta ,\alpha ) \\
&=&\lim_{n\rightarrow 0}\left( \frac{n}{n+1-\beta }\frac{\mathcal{B}%
_{0}^{\left( \beta \right) }\left( \frac{\theta }{\alpha }\right) x^{1-\beta
}}{\alpha ^{\beta }}+\frac{n\mathcal{B}_{1}^{\left( \beta \right) }\left( 
\frac{\theta }{\alpha }\right) x^{2-\beta }}{\alpha ^{\beta }(n+2-\beta )}+%
\frac{n}{\alpha ^{\beta }}\sum\limits_{c=2}^{\infty }\frac{\mathcal{B}%
_{c}^{\left( \beta \right) }\left( \frac{\theta }{\alpha }\right) }{%
(n+c+1-\beta )c!}x^{c+1-\beta }\right) =0,
\end{eqnarray*}%
\begin{eqnarray*}
&&\lim_{x\rightarrow 0}S_{n}(x;\beta ,\theta ,\alpha ) \\
&=&\lim_{x\rightarrow 0}\left( \frac{n}{n+1-\beta }\frac{\mathcal{B}%
_{0}^{\left( \beta \right) }\left( \frac{\theta }{\alpha }\right) x^{1-\beta
}}{\alpha ^{\beta }}+\frac{n\mathcal{B}_{1}^{\left( \beta \right) }\left( 
\frac{\theta }{\alpha }\right) x^{2-\beta }}{\alpha ^{\beta }(n+2-\beta )}+%
\frac{n}{\alpha ^{\beta }}\sum\limits_{c=2}^{\infty }\frac{\mathcal{B}%
_{c}^{\left( \beta \right) }\left( \frac{\theta }{\alpha }\right) }{%
(n+c+1-\beta )c!}x^{c+1-\beta }\right) =0,
\end{eqnarray*}%
and%
\begin{eqnarray*}
&&\lim_{n\rightarrow \infty }S_{n}(x;\beta ,\theta ,\alpha ) \\
&=&\lim_{x\rightarrow \infty }\left( \frac{n\mathcal{B}_{0}^{\left( \beta
\right) }\left( \frac{\theta }{\alpha }\right) x^{1-\beta }}{\alpha ^{\beta
}\left( n+1-\beta \right) }+\frac{n\mathcal{B}_{1}^{\left( \beta \right)
}\left( \frac{\theta }{\alpha }\right) x^{2-\beta }}{\alpha ^{\beta
}(n+2-\beta )}+\frac{n}{\alpha ^{\beta }}\sum\limits_{c=2}^{\infty }\frac{%
\mathcal{B}_{c}^{\left( \beta \right) }\left( \frac{\theta }{\alpha }\right) 
}{(n+c+1-\beta )c!}x^{c+1-\beta }\right) \\
&=&\frac{\mathcal{B}_{0}^{\left( \beta \right) }\left( \frac{\theta }{\alpha 
}\right) x^{1-\beta }}{\alpha ^{\beta }}+\frac{\mathcal{B}_{1}^{\left( \beta
\right) }\left( \frac{\theta }{\alpha }\right) x^{2-\beta }}{\alpha ^{\beta }%
}+\frac{1}{\alpha ^{\beta }}\sum\limits_{c=2}^{\infty }\frac{\mathcal{B}%
_{c}^{\left( \beta \right) }\left( \frac{\theta }{\alpha }\right) }{c!}%
x^{c+1-\beta } \\
&=&\frac{\mathcal{B}_{1}^{\left( \beta \right) }\left( \frac{\theta }{\alpha 
}\right) x^{2-\beta }}{\alpha ^{\beta }}+\frac{1}{\alpha ^{\beta }}%
\sum\limits_{c=2}^{\infty }\frac{\mathcal{B}_{c}^{\left( \beta \right)
}\left( \frac{\theta }{\alpha }\right) }{c!}x^{c+1-\beta }.
\end{eqnarray*}%
For $\alpha =\theta =1$ with $\left\vert t\right\vert <2\pi $, (\ref{1bl})
reduces to the generating function for the Bernoulli numbers of order $\beta 
$. By using same computation in (\ref{1ABN}), we have%
\begin{eqnarray*}
S_{n}(x;\beta ,1,1) &=&\frac{n}{x^{n}}\sum\limits_{c=0}^{\infty }\frac{%
B_{c}^{\left( \beta \right) }}{c!}\int\limits_{0}^{x}t^{n+c-\beta }dt \\
&=&n\sum\limits_{c=0}^{\infty }\frac{B_{c}^{\left( \beta \right) }}{%
(n+c+1-\beta )c!}x^{c+1-\beta },
\end{eqnarray*}%
which yields%
\begin{equation*}
\lim_{n\rightarrow 0}S_{n}(x;\beta ,1,1)=\lim_{n\rightarrow 0}\left( \frac{%
nB_{0}^{\left( \beta \right) }x^{1-\beta }}{n+1-\beta }+n\sum\limits_{c=1}^{%
\infty }\frac{B_{c}^{\left( \beta \right) }}{(n+c+1-\beta )c!}x^{c+1-\beta
}\right) =0,
\end{equation*}%
\begin{equation*}
\lim_{x\rightarrow 0}S_{n}(x;\beta ,1,1)=\lim_{x\rightarrow 0}\left( \frac{%
nB_{0}^{\left( \beta \right) }x^{1-\beta }}{n+1-\beta }+n\sum\limits_{c=1}^{%
\infty }\frac{B_{c}^{\left( \beta \right) }}{(n+c+1-\beta )c!}x^{c+1-\beta
}\right) =0,
\end{equation*}%
and also%
\begin{eqnarray*}
\lim_{n\rightarrow \infty }S_{n}(x;\beta ,1,1) &=&\lim_{n\rightarrow \infty
}\left( \frac{nB_{0}^{\left( \beta \right) }x^{1-\beta }}{n+1-\beta }%
+n\sum\limits_{c=1}^{\infty }\frac{B_{c}^{\left( \beta \right) }}{%
(n+c+1-\beta )c!}x^{c+1-\beta }\right) \\
&=&B_{0}^{\left( \beta \right) }x^{1-\beta }+\sum\limits_{c=1}^{\infty }%
\frac{B_{c}^{\left( \beta \right) }}{c!}x^{c+1-\beta }.
\end{eqnarray*}%
and%
\begin{equation*}
\lim_{x\rightarrow \infty }S_{n}(x;\beta ,1,1)=\frac{B_{\beta -1}^{\left(
\beta \right) }}{(\beta -1)!}
\end{equation*}%
where $\beta \geq 1$. Since%
\begin{equation*}
B_{\beta -1}^{\left( \beta \right) }=(-1)^{\beta -1}(\beta -1)!,
\end{equation*}%
see \cite{Carlitz}, we get%
\begin{equation}
\lim_{x\rightarrow \infty }S_{n}(x;\beta ,1,1)=(-1)^{\beta -1}.  \label{c1}
\end{equation}

For $\beta =1$, the above equation reduces to the generating function for
the Bernoulli numbers. That is%
\begin{equation*}
S_{n}(x;1,1,1)=B_{0}+n\left( \frac{xB_{1}}{n+1}+\frac{B_{2}}{2(n+2)}%
x^{2}+\sum\limits_{c=3}^{\infty }\frac{B_{c}}{(n+c)c!}x^{c}\right) ,
\end{equation*}%
which yields%
\begin{equation*}
\lim_{x\rightarrow 0}S_{n}(x;1,1,1)=\lim_{n\rightarrow
0}S_{n}(x;1,1,1)=B_{0}=1
\end{equation*}%
and%
\begin{equation*}
\lim_{n\rightarrow \infty
}S_{n}(x;1,1,1)=B_{0}+xB_{1}+\sum\limits_{c=2}^{\infty }\frac{B_{c}}{c!}%
x^{c}.
\end{equation*}%
Substituting $\beta =1$ into equation (\ref{c1}), we get%
\begin{equation*}
\lim_{x\rightarrow \infty }S_{n}(x;1,1,1)=1.
\end{equation*}

Assuming that $\left\vert \frac{\alpha }{\theta }e^{-t}\right\vert <1$, with
the aid of binomial series, thus equation (\ref{lc}) reduces to%
\begin{equation*}
\lim_{x\rightarrow \infty }\frac{x^{n}S_{n}(x;\beta ,\theta ,\alpha )}{n}%
=\sum\limits_{j=0}^{\infty }\binom{\beta +j-1}{j}\frac{\alpha ^{j}}{\theta
^{j+\beta }}\int\limits_{0}^{\infty }t^{n}e^{-t(j+\beta )}dt.
\end{equation*}%
Therefore%
\begin{equation*}
\lim_{x\rightarrow \infty }\frac{x^{n}S_{n}(x;\beta ,\theta ,\alpha )}{n}%
=\sum\limits_{j=0}^{\infty }\binom{\beta +j-1}{j}\frac{\alpha ^{j}}{\theta
^{j+\beta }}\frac{1}{(j+\beta )^{n+1}}\int\limits_{0}^{\infty }u^{n}e^{-u}du
\end{equation*}%
By the aid of the Euler gamma function, we get%
\begin{equation}
\lim_{x\rightarrow \infty }\frac{x^{n}S_{n}(x;\beta ,\theta ,\alpha )}{n}%
=\Gamma (n+1)\sum\limits_{j=0}^{\infty }\binom{\beta +j-1}{j}\frac{\alpha
^{j}}{\theta ^{j+\beta }}\frac{1}{(j+\beta )^{n+1}}.  \label{H-1}
\end{equation}%
Assuming that $\beta \in \mathbb{C}\backslash \left\{ 0,-1,-2,-3,\ldots
\right\} $, $n+1\in \mathbb{C}$ when $\left\vert \frac{\alpha }{\theta }%
\right\vert <1\in \mathbb{C};$ $\operatorname{Re}(n+1)>1$ when $\left\vert \frac{%
\alpha }{\theta }\right\vert =1$, then we have%
\begin{equation}
\lim_{x\rightarrow \infty }\frac{x^{n}S_{n}(x;\beta ,\theta ,\alpha )}{n}=%
\frac{\Gamma (n+1)}{\theta ^{\beta }}\Phi _{\beta }\left( \frac{\alpha }{%
\theta },n+1;\beta \right) ,  \label{H2}
\end{equation}%
where $\Phi _{\beta }\left( \frac{\alpha }{\theta },n+1;\beta \right) $
denotes the multiple Hurwitz--Lerch Zeta function, defined by%
\begin{equation*}
\Phi _{n}\left( z,s;a\right) =\sum\limits_{j=0}^{\infty }\binom{n+j-1}{n-1}%
\frac{z^{j}}{(j+a)^{s}}
\end{equation*}%
$a\in \mathbb{C}\backslash \left\{ 0,-1,-2,-3,\ldots \right\} $, $s\in 
\mathbb{C}$ when $\left\vert z\right\vert <1\in \mathbb{C};$ $\operatorname{Re}(s)>1$
when $\left\vert z\right\vert =1$ (cf. \cite[Eq. 16, p. 205]{SrivatavaChoi}).

Substituting the following known formula, given by \cite[Eq. 20, p. 152]%
{SrivatavaChoi},%
\begin{equation*}
\binom{\beta +j-1}{j}=\frac{1}{(j-1)!}\sum%
\limits_{d=0}^{j-1}(-1)^{d+j}S_{1}(j,d+1)\beta ^{d},
\end{equation*}%
where $S_{1}(n,k)$ denotes the Stirling numbers of the first kind%
\begin{equation}
\frac{\left( \log (1+w)\right) ^{k}}{k!}=\sum_{n=0}^{\infty }S_{1}(n,k)\frac{%
w^{n}}{n!}  \label{st}
\end{equation}%
into (\ref{H-1}), we get the following formula:%
\begin{equation*}
\lim_{x\rightarrow \infty }\frac{x^{n}S_{n}(x;\beta ,\theta ,\alpha )}{n}=%
\frac{1}{\theta ^{\beta }\beta ^{n+1}}+\sum\limits_{j=1}^{\infty
}\sum\limits_{d=0}^{j-1}(-1)^{d+j}\frac{\alpha ^{j}\beta ^{d}S_{1}(j,d+1)}{%
(j-1)!\theta ^{j+\beta }(j+\beta )^{n+1}}.
\end{equation*}

Putting $n=1,2,3,4,5$ and $\beta =\theta =\alpha =1$ into (\ref{H-1}), we get%
\begin{equation*}
\lim_{x\rightarrow \infty }xS_{1}(x;1,1,1)=\sum\limits_{j=0}^{\infty }\frac{1%
}{(j+1)^{2}}=\frac{\pi ^{2}}{6},
\end{equation*}%
\begin{equation*}
\lim_{x\rightarrow \infty }\frac{x^{2}S_{1}(x;1,1,1)}{2}=\Gamma
(2)\sum\limits_{j=0}^{\infty }\frac{1}{(j+1)^{3}}=\zeta (3),
\end{equation*}%
\begin{equation*}
\lim_{x\rightarrow \infty }\frac{x^{3}S_{1}(x;1,1,1)}{3}=\Gamma
(3)\sum\limits_{j=0}^{\infty }\frac{1}{(j+1)^{4}}=\frac{\pi ^{4}}{45}
\end{equation*}%
which known as $2\lim_{x\rightarrow \infty }\frac{x^{3}S_{1}(x;1,1,1)}{3}$
is the \textit{crucial for the Debye heat capacity law}.%
\begin{equation*}
\lim_{x\rightarrow \infty }\frac{x^{4}S_{1}(x;1,1,1)}{4}=\Gamma
(5)\sum\limits_{j=0}^{\infty }\frac{1}{(j+1)^{5}}=4!\zeta (5),
\end{equation*}%
\begin{equation*}
\lim_{x\rightarrow \infty }\frac{x^{5}S_{1}(x;1,1,1)}{5}=\Gamma
(6)\sum\limits_{j=0}^{\infty }\frac{1}{(j+1)^{6}}=\frac{5!\pi ^{6}}{945},
\end{equation*}%
and so on.

Consequently, when $x\rightarrow \infty $, the asymptotic behavior of the
function $S_{n}(x;\beta ,\theta ,\alpha )$ is given as follows:

\begin{theorem}
Let $n\in \mathbb{N}$ and $\beta >0$. Then we have%
\begin{equation}
S_{n}(x;\beta ,\theta ,\alpha )\approx \frac{n\Gamma (n+1)}{x^{n}\theta
^{\beta }}\Phi _{\beta }\left( \frac{\alpha }{\theta },n+1;\beta \right) .
\label{ap}
\end{equation}
\end{theorem}

Substituting $\beta =\theta =\alpha =1$ into (\ref{ap}) and $n\in \mathbb{N}$%
, we have the following results:

\begin{equation*}
S_{2n+1}(x;1,1,1)\approx \frac{2n+1}{x^{2n+1}}\Gamma (2n+2)\zeta \left(
2n+2\right)
\end{equation*}%
Since%
\begin{equation*}
\zeta \left( 2n+2\right) =\frac{(-1)^{n}\left( 2\pi \right) ^{2n+2}}{2(2n+2)!%
}B_{2n+2}
\end{equation*}%
(cf. \cite{SrivatavaChoi}) and%
\begin{equation*}
\Gamma (2n+2)=(2n+1)!,
\end{equation*}%
we get the following corollary:

\begin{corollary}
Let $n\in \mathbb{N}$. Then%
\begin{equation*}
S_{2n+1}(x;1,1,1)\approx \frac{(-1)^{n}\left( 2\pi \right) ^{2n+2}\left(
2n+1\right) (n+1)}{x^{2n+1}}B_{2n+2}.
\end{equation*}%
Thus, it is easy to see that%
\begin{equation*}
d_{n}(x):=S_{n}(x;1,1,1)\approx \frac{n}{x^{n}}\Gamma (n+1)\zeta \left(
n+1\right) ,
\end{equation*}%
see also \cite{Engeln}. Therefore, due to (\ref{dB}), it is clear to see that%
\begin{equation*}
\lim_{x\rightarrow 0}d_{n}(x)=1
\end{equation*}%
(\textit{cf}. \cite{Abro}, \cite{Cox}, \cite{Debey}, \cite{Engeln}, \cite%
{Epap}, \cite{Gonzalez}, \cite{Grad}, \cite{Guagliardi}, \cite{Guagliardi}, 
\cite{Guseinov}, \cite{Hall}, \cite{Harrison}, \cite{Jordan}, \cite%
{Khishchenko}, \cite{Macleod}, \cite{Ng}, \cite{Passler}).
\end{corollary}

\section{Conclusion}

For $n$, $\alpha $, $\beta $, and $\theta $ are real or complex parameters,
we gave some properties of the following function:%
\begin{equation*}
f_{n}(t;\beta ,\theta ,\alpha )=\frac{t^{n}}{\left( \theta e^{t}-\alpha
\right) ^{\beta }}.
\end{equation*}%
We showed that this function related to generating functions, Moment
generating function for the negative binom distribution, the
Apostol-Bernoulli numbers, the multiple Hurwitz Lerch zeta functions, the
Riemann zeta function, the Frobenius-Euler numbers, and the Bernoulli
numbers, etc. We defined unified the Debye function for special values of
the parameters $\beta ,\theta ,\alpha $:%
\begin{equation*}
S_{n}(x;\beta ,\theta ,\alpha )=\frac{n}{x^{n}}\int\limits_{0}^{x}f_{n}(t;%
\beta ,\theta ,\alpha )dt,
\end{equation*}%
where $\left\vert t\right\vert <2\pi $ when $\alpha =\theta =1$; $\left\vert
t\right\vert <\left\vert \ln \left( \frac{\theta }{\alpha }\right)
\right\vert $ when $\frac{\theta }{\alpha }\neq 1$\ and $\operatorname{Re}(n)\geq 1$%
. $\ln w$\ ($\left\vert \arg (w)\right\vert <1$) denotes the principal
branch of the many-valued function $\ln (w)=\ln \left\vert w\right\vert
+i\arg (w)$ ($\left\vert w\right\vert >0$). We gave a recurrence relation
and Maclaurin's series expansion of the function $S_{n}(x;\beta ,\theta
,\alpha )$. By using some special values of the Maclaurin's series expansion
of the function $S_{n}(x;\beta ,\theta ,\alpha )$, we also showed that the
function $S_{n}(x;\beta ,\theta ,\alpha )$ related to many special numbers
and polynomials, multiple Hurwitz Lerch zeta function, Riemann zeta
function, the Apostol-Bernoulli numbers and the Bernoulli numbers. Finally,
we gave approximation formula with the aid of integral and series
presentations for the function $S_{n}(x;\beta ,\theta ,\alpha )$. That is,
for $n\in \mathbb{N}$ and $\beta >0$, we prove that%
\begin{equation*}
S_{n}(x;\beta ,\theta ,\alpha )\approx \frac{n\Gamma (n+1)}{x^{n}\theta
^{\beta }}\Phi _{\beta }\left( \frac{\alpha }{\theta },n+1;\beta \right) .
\end{equation*}%
The results of this paper will be potentially used in mathematics, physic
and chemistry.

\end{document}